\documentclass[12pt]{article}
\usepackage{amsmath,amssymb,exscale}
\usepackage{graphicx}
\usepackage{epsfig}
\usepackage{multicol}
\usepackage{color}
\usepackage{hyperref}
\hypersetup{colorlinks,bookmarksopen,bookmarksnumbered,citecolor=rossos,
linkcolor=black,pdfstartview=FitH,urlcolor=blus}
\usepackage{slashed}

\definecolor{blus}{cmyk}{1,1,0,0.6}
\definecolor{verdes}{cmyk}{0.99,0,0.59,0.82}
\definecolor{rossos}{cmyk}{0,1,1,0.55}
\definecolor{greeny}{cmyk}{0.99,0,0.59,0.98}

\newcommand{\tmtextbf}[1]{{\bfseries{#1}}}
\newcommand{\tmtextrm}[1]{{\rmfamily{#1}}}
\def\be{\begin{equation}}
\def\ee{\end{equation}}
\def\bea{\begin{eqnarray}}
\def\eea{\end{eqnarray}}

\definecolor{red}{rgb}{1,0,0}

\newcommand{\GL}{{\scriptscriptstyle\rm GL}}
\newcommand{\gappeq}{{\rlap{{\raise}.5ex\text{\ensuremath{>}}}{{\lower}.5ex\text{\ensuremath{\sim}}}}}
\newcommand{\lappeq}{{\rlap{{\raise}.5ex\text{\ensuremath{<}}}{{\lower}.5ex\text{\ensuremath{\sim}}}}}

\newcommand{\I}{\tmtextrm{1{\kern}-.24em l}}

\begin{document}
\topmargin -1.0cm
\oddsidemargin -0.5cm
\evensidemargin -0.5cm

{\vspace{-1cm}}
\begin{center}

\vspace{-1cm}

 {\LARGE \tmtextbf{ 
{\color{verdes}Holographic Superfluids and Superconductors \\ in Dilaton-Gravity }}} {\vspace{.5cm}}\\

\vspace{0.5cm}

{\large  Alberto Salvio 
%\vspace{.4cm}\\
%{\large }\\
\vspace{.3cm}

 {\it Scuola Normale Superiore and INFN, Piazza dei Cavalieri 7, 56126 Pisa, Italy}

\vspace{0.5cm}

}
%\vspace{.4cm}
 \end{center}
 \begin{abstract}

\noindent We investigate holographic models of superfluids and superconductors in which the gravitational theory includes a dilatonic field. Dilaton extensions are interesting as they allow us to obtain 
a better description of low temperature condensed matter systems. We focus on asymptotically AdS black hole configurations, which are dual to field theories with conformal ultraviolet behavior.
A nonvanishing value of the dilaton breaks scale invariance in the infrared and is therefore compatible with the normal phase being insulating (or a solid in the fluid mechanical interpretation); indeed we find that 
this is the case at low temperatures  and if one appropriately chooses the parameters of the model. Not only the superfluid phase transitions, 
but also the response to external gauge fields is analyzed. This allows us to study, among other things, the vortex phase and to show that these holographic superconductors are also of Type II.
However, at low temperatures they can behave in a qualitatively different way compared to their analogues without the dilaton: the critical magnetic fields and the penetration depth can remain finite in the small 
$T/T_c$ limit.

\end{abstract}

\tableofcontents

%\newpage

\section{Introduction and results} \label{introduction}

The anti de Sitter/conformal field theory (AdS/CFT) correspondence is a powerful method to investigate strongly coupled theories. An interesting application of such duality is the study of systems 
at finite temperature and density of some conserved quantum number.
This is the typical situation in condensed matter physics and in the last few years a considerable amount of work has been dedicated to the development of holographic methods for condensed matter systems (reviews of which can be found in 
Refs. \cite{Hartnoll:2009sz,McGreevy:2009xe,Hartnoll:2009qx,Sachdev:2010ch,Pires:2010mt,Horowitz:2010nh,Hartnoll:2011fn}).  

In the simplest realizations one deals with Einstein's gravitational theory with a negative cosmological constant coupled to a gauge field (dual to the conserved current). 
While this setup captures some interesting properties of realistic systems, 
the low temperature behavior has some disturbing features. A classic example, which is perhaps unphysical rather than disturbing, is the fact that the simplest charged black hole solution predicts a nonvanishing
 entropy in the zero temperature limit. This situation can be improved \cite{Gubser:2009qt,Goldstein:2009cv,Charmousis:2010zz} by adding a real scalar field $\phi$ ({\it a dilaton}), which interacts with the gauge field 
by modifying its gauge coupling $g$,
\be \frac{1}{g^2} \rightarrow \frac{1}{g^2(\phi)} = \frac{Z_A(\phi)}{g^2}, \label{running-coupling} \ee
where $Z_A\neq 0$, and self-interacts  through a
potential. This allows for zero entropy at zero temperature. Moreover, real scalars of this type are typically present in the low energy limit of string theories and their inclusion may be good if eventually one has to find a stringy embedding. 

In this work we initiate the study of holographic superfluids and superconductors in dilaton extensions of Einstein's gravity. 
Like in the absence of the dilaton, the minimal description of these systems requires a conserved U(1) current and a charged condensing operator, which are respectively dual 
to an Abelian gauge field and a charged scalar living on an asymptotically AdS space \cite{Gubser:2008px,Hartnoll:2008vx,Herzog:2008he,Hartnoll:2008kx} 
(for a review see \cite{Hartnoll:2009sz,Herzog:2009xv,Horowitz:2010gk,Horowitz:2010nh}). Then the dilaton  can couple directly not only to the gauge field but also to the charged scalar\footnote{Our setup is different from that in Ref. \cite{Liu:2010ka} because the dilaton
there is defined as the modulus of the charged scalar rather than a real scalar, as it is usually done.}. 
In the case of superconducting materials the U(1) symmetry is gauged while for superfluids is global. In holography a superfluid corresponds to a Dirichlet condition for the gauge field at the AdS boundary, 
while a true superconductor can be obtained by imposing  a boundary condition
 of the Neumann type \cite{Domenech:2010nf}. Superfluids can be considered as superconductors in the limit of nondynamical (electromagnetic)
gauge fields. For the sake of definiteness, we adopt the terminology used in the literature of superconductivity, rather than that of superfluidity.

One of the main motivations for  holographic superconductivity is the hope to shed light on high-$T$ superconductors. Experiments indeed imply that these
materials cannot be entirely described by the weakly coupled theory of Bardeen, Cooper, and Schrieffer (BCS), but they could involve strong coupling.
 The cuprate high-$T$ superconductors can be obtained by doping a Mott insulator and in the ($T$ versus ``dopant concentration'') phase diagram  one can observe 
an insulator/superconductor transition in the low temperature region \cite{doping}; therefore the normal (nonsuperfluid) phase is expected to  be gapped.
 A transition of this sort has been  realized in holography  by considering a theory with a
compactified  extra dimension \cite{Nishioka:2009zj,Horowitz:2010jq} (see also \cite{Montull:2012fy} for the realization of a true superconductor with a dynamical gauge field). The size of this extra dimension breaks scale invariance and at low temperatures
 leads to a gap of charged excitations and thus to an insulator normal phase. One expects to be able to obtain an insulator/sperconductor transition by introducing a dilaton, even in the absence of 
additional dimensions, because  a running dilaton also introduces a scale. We will find that this is indeed the case at low enough temperatures 
and for some values of the parameters.

Let us describe in more detail our main results and the organization of the paper. In section \ref{model} we will define our Einstein-dilaton model coupled to the U(1) gauge field and the charged scalar and recall the AdS/CFT dictionary. 
Here and in the rest of the paper we focus on 2+1 dimensional
CFTs. In the same section we will also review the (recently found \cite{Anabalon:2012ta}) most general static and asymptotically AdS planar black hole of the dilaton-gravity system. 

Once we fix the parameters of the model, there is only one solution with active dilaton (dilaton-BH), but the fact that the space is asymptotically AdS implies that the Schwarzschild black hole (S-BH), which has $\phi=0$, is also a solution.
 Therefore, in section \ref{Einstein-dilaton section}, we 
compare their free energies as a function of $T$ to determine which one is  energetically favorable. We find that the S-BH is favorable at high temperatures, while the dilaton-BH dominates in the 
low temperature region. In the latter phase the dilaton runs logarithmically with the energy and grows  from the ultraviolet (UV) to the infrared (IR).
 We then study the AC and DC conductivity by considering an electromagnetic (EM) perturbation on top of these geometries. The presence of a black hole horizon implies that 
the system is never perfectly insulating, but for some choices of $Z_A(\phi)$ % (those that are small at the black hole horizon) 
the DC conductivity is suppressed especially at low temperatures, resembling an insulator. This occurs when the running coupling constant $g^2(\phi)$ becomes large in the IR and, not surprisingly, links 
the insulating phase with IR strong coupling.

In section \ref{homogeneousSFphase} we find superfluid phase transitions by solving the  gauge field and  charged scalar equations on the fixed dilaton-BH geometry. Like for the S-BH \cite{Hartnoll:2008vx}, for each value of the temperature and the other parameters 
there is a critical temperature $T_c$ below which the charged operator condenses breaking the U(1) symmetry. We also study the conductivity in the broken phase and obtain that both the gap of charged excitations and the superfluid 
density are modified by changing the couplings of the dilaton, in particular the running constant, Eq. (\ref{running-coupling}).

Some of the most famous properties of superconductors occur in the presence of external magnetic fields; 
therefore we move on and study the magnetic response in the dilaton-BH phase. The same type of analysis in the model 
without the dilaton has been performed in \cite{Hartnoll:2008kx,Albash:2008eh,Domenech:2010nf,Montull:2012fy}. While for superfluids the external and total
magnetic fields coincide, they differ substantially in true superconductors. In section \ref{HSCs section} we study the superconductor case. We directly observe the Meissner effect and find vortex solutions, 
the penetration depth and the critical magnetic fields $H_{c1}$ and $H_{c2}$. Like in the case with no dilaton \cite{Domenech:2010nf,Montull:2012fy}, we always find $H_{c1}<H_{c2}$ (the superconductor is of Type II) and thus this seems to be 
a rather general prediction of holography. Quite interestingly, the known high-$T$ superconductors are of this type.
Crucial differences with respect to the S-BH phase \cite{Domenech:2010nf} emerge in the low temperature region, for some choices of $Z_A(\phi)$ (which, interestingly, lead to an insulating normal phase). In these cases the ratio $H_{c2}/H_{c1}$ remains finite in the limit of small $T/T_c$
unlike the S-BH case \cite{Domenech:2010nf} and also the penetration depth shows an analogue difference. Finally in section \ref{superlfuids} we study superfluids in the presence of external rotations and find 
superfluid vortices in the dilaton-BH phase; here we find results that are qualitatively similar to those for the S-BH \cite{Montull:2009fe,Keranen:2009re,Domenech:2010nf}, although the dilaton generically modify the vortex shape.

Section \ref{conclusions-section} contains some outlook of our work.

\section{The model and the AdS/CFT correspondence}\label{model}

Our model is defined by the usual Einstein-dilaton action coupled to a U(1) gauge field $A_{\alpha}$ and a charged scalar $\Psi$:  \be S=\int d^{4}x\, \sqrt{-g}\left\{{{1\over16\pi
G_N}}\left[\mathcal{R}-(\partial_{\alpha} \phi)^2 - V(\phi) \right]\,-\frac{Z_{A}(\phi)}{4g^2}{\cal F}_{\alpha \beta}^2-\frac{Z_{\psi}(\phi)}{L^2g^2}|D_\alpha\Psi|^2\right\}, \label{action} \ee 
where $\mathcal{R}$ is the Ricci scalar, $G_N$ 
the gravitational Newton constant, $\phi$ a real scalar (the dilaton)
and  we introduced ${\cal F}_{\alpha \beta}= \partial_\alpha A_\beta-\partial_\beta A_\alpha$  and $D_{\alpha}=\partial_{\alpha}-iA_{\alpha}$. A negative cosmological constant $\Lambda$ is included in the dilaton potential $V(\phi)$, and 
the AdS radius is defined by $\Lambda=-6/L^2$. We define $\phi$ such that $V(0)= \Lambda$. Also the dilaton couples to $A_{\alpha}$ and $\Psi$ through two 
generic functions $Z_{A}(\phi)$ and  $Z_{\psi}(\phi)$ and we choose the normalization of the matter fields in a way that  $Z_{A}(0)=Z_{\psi}(0)=1$. There are no special requirements for the $Zs$ at this level, besides the fact
that they should be regular and nonvanishing for any $\phi$ in order for the semiclassical approximation to be valid.  For simplicity,    we have not added any
potential for the charged scalar.

The most general static asymptotically AdS planar black hole with two-dimensional rotation and translation invariance has the form
\be ds^2=W(z)\left(-f(z)dt^2+dy^2+\frac{dz^2}{f(z)}\right)  ,\
\ \phi=\phi(z), \label{generalBH} \ee 
where $t$ represents time, $dy^2=\delta_{ij}dy^idy^j$ and $z$ is the holographic coordinate, which we define such that the AdS-boundary is located at $z=0$: for $z$ close to zero, $W$, $f$ and $\phi$
approximate to $W(z)=L^2/z^2$, $f(z)=1$ and $\phi(z)=0$. This allows us to use the standard AdS/CFT dictionary, which relates the property of a gravitational theory with those of a CFT. In particular the boundary 
values of $\Psi$ and $A_{\mu}$ are the sources of a charged operator $\mathcal{O}$ and the current $J_{\mu}$ in the CFT side. Therefore, if one solves the classical field equations with boundary conditions 
\be \Psi|_{z=0}=\Psi_0, \qquad A_{\mu}|_{z=0}=a_{\mu}, \label{s-amu}\ee
 the Green's function for $\mathcal{O}$ and $J_{\mu}$ are given by functional derivatives of the on-shell action with respect to $\Psi_0$ and $a_{\mu}$. For example the vacuum expectation values are given by
\be 
\langle  J_\mu\rangle=\frac{1}{g^2}{\cal F}_{z\mu}|_{z=0}\ , \ \  \langle{\cal
O}\rangle=\frac{1}{g^2 z^{2}}D_z \Psi|_{z=0}\, .
\label{operator} \ee

We will work in the limit  $G_N\rightarrow0$
 such that  the effect of
the gauge field and the charged scalar on the metric and the dilaton can be neglected.
In this limit the most general asymptotically AdS  black hole of the form given in Eq. (\ref{generalBH}) is \cite{Anabalon:2012ta}  
\bea W(z)=\frac{\nu^2\left(1+z/L\right)^{\nu-1}}{\left[\left(1+z/L\right)^{\nu}-1\right]^2}, \
\ \phi(z)= \pm \sqrt{\frac{\nu^2-1}{2}}\ln(1+z/L), \nonumber \\ f(z)=1 + 3\left(\frac{L}{z_s}\right)^3\left\{\frac{\nu^2}{4-\nu^2}+(1+z/L)^2\left[1-\frac{(1+z/L)^{\nu}}{\nu+2}-\frac{(1+z/L)^{-\nu}}{2-\nu}\right]\right\}. 
\label{dilaton-BH}\eea 
 The black hole horizon $z_h$, defined as usual by $f(z_h)=0$, is a function\footnote{By dimensional analysis $z_h/L$ is a function of $\nu$ and $z_s/L$ only.}  of $\nu$ and $z_s$. Since we want to study the theory at finite temperature,
we perform the Euclidean continuation to imaginary time $it$ with period $\beta=1/T$. The temperature can then be computed with the formula $T=|f'(z_h)|/4\pi$, where a prime denotes a derivative with respect to $z$. 

When $\nu=1$ we have $\phi=0$ and we recover  the S-BH: $W(z)=L^2/z^2, f(z)=1-(z/z_s)^3$.  In this case $z_s=z_h$ and we have $z_h=3/(4\pi T)$.
If instead $\nu\neq 1$ the dilaton is active and its backreaction on the metric is nontrivial. This results in a displacement of the horizon and the temperature is no longer given by $3/(4\pi z_s)$, but by a more general ($\nu$-dependent) expression. 
It is important to notice that the S-BH is a solution whenever $V$ has a  negative stationary point. In order to have the dilaton-BH, $\nu\neq 1$, one has to consider instead the following family of potentials  \cite{Anabalon:2012ta}
\be V(\phi)= \sum_{i=1}^{6} V_ie^{-\delta_i \phi},\ee
where $V_i$ and $\delta_i$ are given by
\bea \frac{z_s^3}{L}V_i&=&3  \frac{1-\nu }{2 + \nu}, \,\,\,\, 12\frac{1-\nu^2}{\nu^2-4},\,\,\,\, (1+\nu )\frac{ 3  \nu^2- \left(\nu^2-4\right)z_s^3/L^3}{\nu ^2(\nu -2)}, \nonumber \\ 
&&  3  \frac{\nu+1 }{2 - \nu} , \,\,\,\,
4(\nu^2-1)\frac{ 3  \nu^2- \left(\nu^2-4\right)z_s^3/L^3}{\nu ^2(\nu^2 -4) },  \,\,\,\,(\nu -1 )\frac{ 3  \nu^2- \left(\nu^2-4\right)z_s^3/L^3}{\nu ^2(\nu +2) }, \eea
and 
\be \mp \sqrt{\frac{\nu^2-1}{2}}\,\,\delta_i= \nu +1  , \,\,\,\, 1 , \,\,\,\, \nu-1  , \,\,\,\, 1-\nu   , \,\,\,\, -1 , \,\,\,\, -\nu-1 \ee
respectively, where the $\mp$ signs correspond to $\phi\geq 0$ and $\phi<0$; the potential is therefore symmetric under $\phi\rightarrow -\phi$ and from now on we restrict to $\phi\geq 0$.
 Since the black hole is asymptotically AdS, this potential has a stationary point for $\phi=0$ such that $V(0)=\Lambda$. An immediate but important  consequence is that one has two black holes for any $\nu\neq 1$, the S-BH and the one with active dilaton.
For  $\phi$ small enough one obtains  $V(\phi)\simeq \Lambda -\frac{2}{L^2}\phi^2$.
Notice  that the squared mass of $\phi$ is always above the Breitenlohner-Freedman bound \cite{Breitenlohner:1982jf} and therefore $\phi=0$ is never unstable, but either metastable or stable. 

%here the comment-dilaton-potential.tex should be put

\section{The Einstein-dilaton theory} \label{Einstein-dilaton section}

In this section we focus on the normal phase, $\langle \mathcal{O}\rangle =0$. This, among other things, will help us to understand what kind of superfluid transitions are those described in the rest of the paper.

\subsection{Phases} \label{pahses}

Let us compare the free energies of the S-BH and the dilaton-BH to determine which one is energetically favorable, as a function of $T$.
The free energy $F$ can be computed by evaluating the euclidean action $S_E$ at the solution of interest,
\be F=\beta S_E.\label{free-energy}\ee 
According to this definition it is clear that the free energy of the S-BH is divergent. One reason is that the solution does not depend on the planar coordinates y and therefore 
$F$ goes to infinity as the volume of the plane $V=\int d^2y\rightarrow \infty$. This is an infrared divergence and can be avoided by considering the free energy density $F/V$. 
There is, however, another source of singularity, due to the fact that the Lagrangian evaluated at the S-BH goes as $1/z^4$ for $z$ small enough and the integral includes the point $z=0$; however this is an ultraviolet
divergence and as such it should cancel in the difference $F_d-F_s,$ where $F_d$ and $F_s$ are the free energy densities of the dilaton- and S-BH respectively. Indeed the dilaton-BH is asymptotically AdS and therefore its ultraviolet behavior should coincide with that 
of the S-BH.

\begin{figure}[ht]
  % \begin{tabular}{cc}
   {\hspace{-1.5cm}}
  \hspace{4cm} \includegraphics[scale=1]{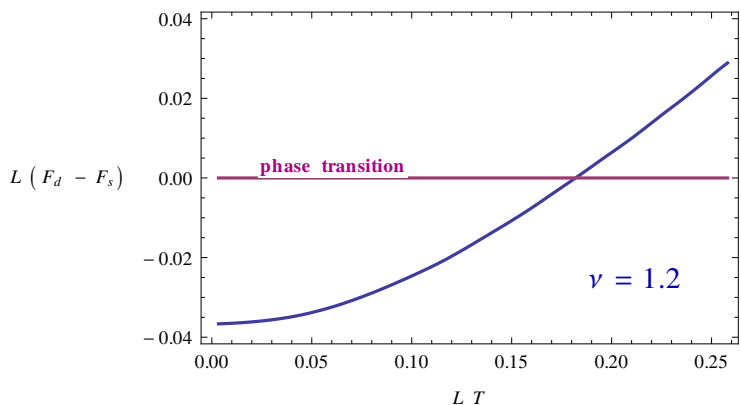}  
 
%      {\hspace{1.cm}} 
   % \includegraphics[scale=0.7]{FvsT.eps}  
%    \end{tabular}
   \caption{{\small {\it The free energy density (up to the gravitational constant $1/16\pi
G_N$) of the dilaton-BH minus that of the S-BH as a function of the temperature.}}}
\label{phase-transition-D-d-BH}
\end{figure}

We plot the free energy difference in Fig. \ref{phase-transition-D-d-BH}. We find that if we lower the temperature below a critical value $\mathcal{T}_c$ there is a phase transition between the S-BH and the dilaton-BH. 
 Since $L$ is the only dimensionful parameter, $\mathcal{T}_c\propto 1/L$, but the actual value of $\mathcal{T}_c$ depends on the dimensionless parameter $\nu$ as well. 

We observe that the entropy, $-\partial_T F$, approximates to zero at small temperatures.

\subsection{Conductivity} \label{Conductivity-section}

Both the S-BH and dilaton-BH phases correspond to systems with nonvanishing conductivity in the CFT side. This property can be viewed essentially as a consequence of the black hole horizon (see below). However, as we will see, the 
DC conductivity can be suppressed for some values of the parameters, resembling an insulator.

 To compute the conductivity let us consider, on top of these geometries, a small time-dependent perturbation along a spatial coordinate $x$, 
\be A_x (t,z)=\mathcal{A}(z)e^{i\omega (p(z)-t)}, \label{wave-decomposition}\ee
 which is induced by a small electromagnetic field at the
 AdS-boundary, $a_x (t)$. Here $\mathcal{A}$ and $p$ are real functions of $z$. The system responds creating a current which is linear in $a_x$: $\langle J_x\rangle= \sigma E_x$, where $\sigma$ is the complex conductivity and $E_x=-\partial_t a_x$ the electric field.  
Using the AdS/CFT dictionary, the first equation in (\ref{operator}), we have
\be g^2\sigma=  p'(0)-i\frac{ \mathcal{A}'(0)}{\omega \mathcal{A}(0)}. \label{sigma}\ee
%
%where a prime denotes a derivative with respect to $z$.
 Since this is a linear response problem the conductivity can  be computed by solving the linearized Maxwell equation
\be \partial_z(f Z_A(\phi) \partial_zA_x)+ \omega^2 \frac{Z_A(\phi)}{f} A_x= 0 \label{linearizedEOM}\ee 
with appropriate boundary conditions. At the AdS-boundary the condition  is, as we said, $A_x|_{z=0}=a_x$,  while at the black hole horizon we impose that the EM wave is going towards the horizon (ingoing boundary condition)
\be p'(z)f(z)\rightarrow 1 \quad \mbox{at } \quad z=z_h, \label{IRBC}\ee 
  as required by regularity \cite{Hartnoll:2009sz}. 

Inserting (\ref{wave-decomposition}) into (\ref{linearizedEOM}) we obtain an equation whose imaginary part can be solved analytically, to obtain
\be p'(z)=\frac{p'(0)}{Z_A(\phi(z))f(z)} \left(\frac{\mathcal{A}(0)}{\mathcal{A}(z)}\right)^2. \label{Imeq}\ee 
A consequence is that $p'(0)$ cannot be zero without violating Eq. (\ref{IRBC}).
Therefore the amplitude of the EM wave, $\mathcal{A}(z)e^{i\omega p(z)}$, acquires an imaginary part which
%\footnote{It is indeed possible to show that there are no solutions to (\ref{linearizedEOM}) satisfying (\ref{IRBC})
%  and having at the same time  
% $p'(z=0)=0$.}
 generates a nonvanishing Re$[\sigma]$. %This means that the dilaton-BH and the S-BH are both metallic phases.
 In the fluid mechanical interpretation Re$[\sigma]\neq 0$ corresponds to a fluid, as opposed 
to a solid which has vanishing Re$[\sigma]$, as discussed in \cite{Montull:2012fy}. 

The real part of Eq. (\ref{linearizedEOM}) is instead
\be \partial_z(fZ_A\partial_zs) +fZ_A (\partial_zs)^2+\omega^2 \frac{Z_A}{f} \left(1-\frac{e^{-4s}}{Z_A^2}\right)=0, \label{eq-s}\ee 
where $$s(z)\equiv \ln\frac{\mathcal{A}(z)}{\mathcal{A}(0)}-\frac{1}{2}\ln p'(0),$$ to be solved with the regularity boundary conditions
\be s=-\frac{1}{2}\ln Z_A, \qquad s'=-\frac{2\omega^2 Z_A'/Z_A}{4\omega^2+f'^2}, \qquad \mbox{at } \quad z=z_h. \label{bc-s}\ee
Once a solution to this equation is found we can compute the real and imaginary part of the conductivity through
\be \mbox{Re}[\sigma]=\frac{e^{-2s(0)}}{g^2} ,\qquad \mbox{Im}[\sigma]=-\frac{s'(0)}{g^2 \omega},  \label{compute-sigma}\ee
where we used Eqs. (\ref{sigma}) and (\ref{Imeq}). 

\begin{figure}[t]
   \begin{tabular}{cc}
   {\hspace{-0.5cm}}
   \includegraphics[scale=0.8]{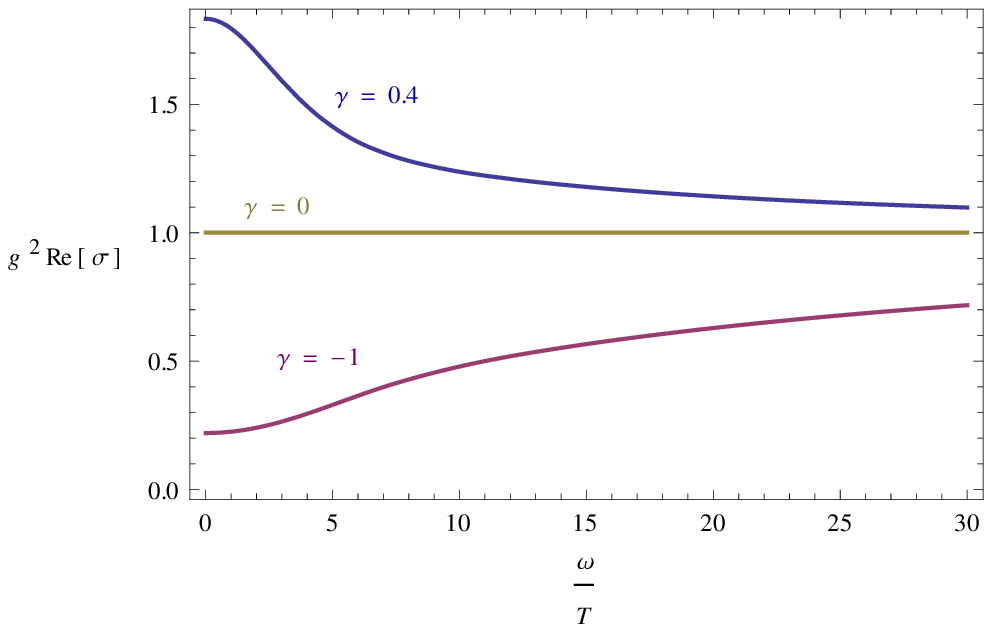}  
 
     {\hspace{1.cm}} 
    \includegraphics[scale=0.82]{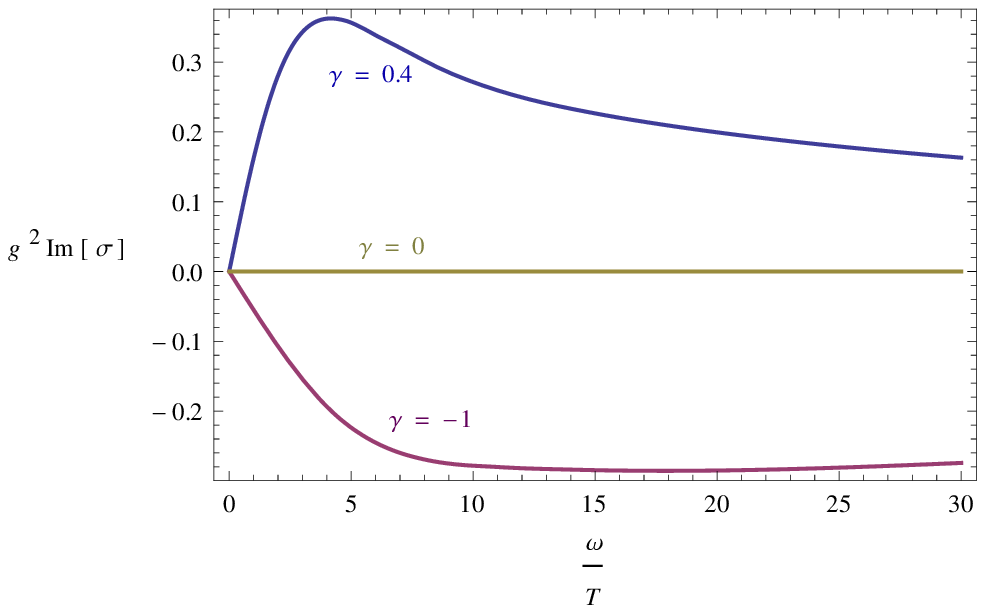}  
    \end{tabular}
   \caption{{\small {\it  The conductivity as a function of the frequency in the normal phase, $\langle \mathcal{O }\rangle=0$. We set $T\simeq 0.010/L$, $\nu=1.2$, $Z_A(\phi)=e^{\gamma\phi}$ and $Z_\psi=1$.}}}
\label{sigma-NP}
\end{figure}

A particularly  simple  case is when there is no direct coupling between the dilaton and the gauge field, $Z_A=1$, which corresponds to $s=0$ and thus to a frequency-independent Re$[\sigma]$ and to 
a vanishing Im$[\sigma]$, both for the S-BH and the dilaton-BH. In this case the dilaton does not have an impact on $\sigma$; it coincides with that found in \cite{Hartnoll:2008vx}.

 When $Z_A$ depends on $\phi$ the situation is different. 
A simple way to see it is to take the zero frequency limit. Then Eqs. (\ref{eq-s}) and (\ref{bc-s}) tell us $s(z)=-\frac{1}{2}$ln$Z_A|_{z=z_h}$ for every $z$, which leads to 
\be \lim _{\omega \rightarrow 0} \mbox{Re}[\sigma]=\frac{1}{g^2}Z_A|_{z=z_h}, \qquad  \lim _{\omega \rightarrow 0}\mbox{Im}[\sigma]=0. \label{DCsigma}\ee
Here we see that the DC conductivity can be suppressed or enhanced depending on the function $Z_A(\phi)$. For the exponential form $Z_A(\phi)=e^{\gamma \phi}$, the more positive $\gamma$ is the bigger the DC conductivity, while a large 
negative value of $\gamma$ corresponds to an approximate insulating behavior. This effect is even stronger at low temperatures where the dilaton is logarithmically large. We have 
\be \lim _{\omega \rightarrow 0} \mbox{Re}[\sigma] \sim T^{-\gamma \sqrt{(\nu^2-1)/2}} \qquad \mbox{at low temperatures},\label{sigma-vs-T}\ee
where we have remembered that we restrict to the positive sign choice in (\ref{dilaton-BH}), without loss of generality.
The conductivity can depend on $T$ because there is an additional scale in the dilaton phase. It is interesting to notice that for the functions $Z_A(\phi)$ such that the DC conductivity is small the phase transition described in Section 
\ref{pahses} is between a conductor at high temperatures and a (non ideal) insulator at low temperatures, while in the other cases we observe a transition between two metallic 
phases; in particular if we set $\gamma=\sqrt{2/(\nu^2-1)}$ in Eq. (\ref{sigma-vs-T}) we get a low temperature resistivity that goes linearly in $T$, like for many unconventional (not BCS) superconductors.

In the  high frequency limit, instead, Eqs. (\ref{eq-s}) and (\ref{bc-s}) tell us that $s\rightarrow -\frac{1}{2}\ln Z_A$ and so  Eqs. (\ref{compute-sigma}) imply that $\sigma$ converges to the constant value we have in the S-BH phase: 
Re$[\sigma]=1/g^2$ and Im$[\sigma]=0$. At high enough energies these systems respond in a universal way. % as they all have a common UV completion.

In Fig. \ref{sigma-NP} we give $\sigma$ as a function of $\omega$. A direct coupling between the dilaton and the bulk gauge field gives us a nontrivial frequency dependence of both Re$[\sigma]$ and Im$[\sigma]$,
unlike the S-BH phase (and the case $Z_A(\phi)=1$).
As suggested by the first equation in (\ref{DCsigma}), Re$[\sigma(\omega)]$ is large (small) when $Z_A(\phi)$ is large (small) at the horizon. We observe that the same is true for Im$[\sigma]$. However, while for Im$[\sigma]$ the effect of the 
dilaton coupling function is strongest at intermediate values of  $\omega$ (as implied by the analytic discussion above), for Re$[\sigma]$ we observe the maximum effect at low frequencies.

\section{Superfluid phase transitions} \label{homogeneousSFphase}

In this section we obtain superfluid phase transitions by introducing a chemical potential. At low temperatures, for the functions $Z_A(\phi)$ such that the normal phase resembles an insulator, these are 
new insulator/superconductor transitions.

\subsection{Homogeneous superconducting phase}

In order to obtain superconducting solutions the simplest ansatz is 
\be \Psi=\psi(z),\quad A_0=A_0(z) ,\label{ansatz} \ee
where $\psi$ is a real function, and the other components of $A_{\alpha}$  are set equal to zero. A nonvanishing value of $A_0$ is needed 
in order to have a regular solution for the condensate profile $\psi(z)$ (see below). Therefore we consider the following AdS-boundary conditions (see Eq. (\ref{s-amu}))
\be \Psi_0=0, \qquad a_0=\mu. \label{UV}\ee
The first condition ensures that the U(1) breaking is spontaneous, while the chemical potential $\mu$ is needed to have $A_0\neq 0$. The regularity at the black hole horizon
indeed requires $A_0(z_h)= 0$ (in order for $A_0$ to have finite norm) and therefore $a_0=0$ would imply $A_0=0$ everywhere.

\begin{figure}[t]
  % \begin{tabular}{cc}
   % {\hspace{-0.5cm}}
  {\hspace{2cm}}    \includegraphics[scale=1.1]{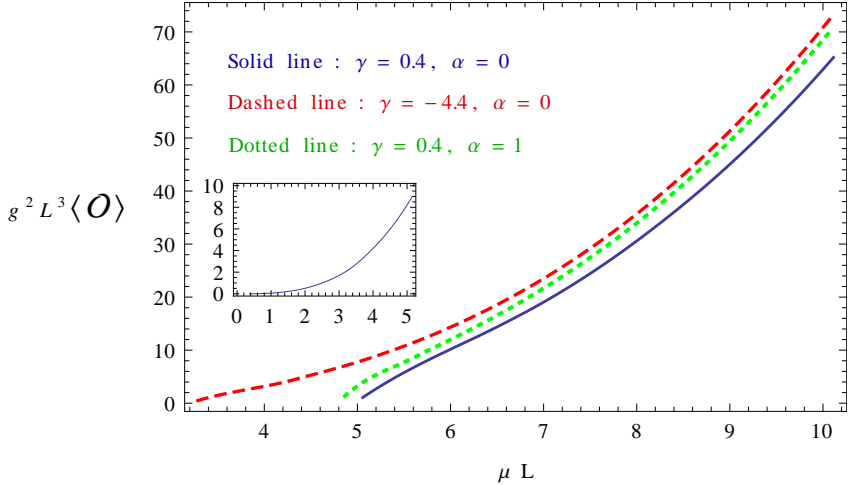}  
 
   %  {\hspace{1.cm}} 
 %   \includegraphics[scale=0.7]{Ovs-mu2.eps}  
%    \end{tabular}
   \caption{{\small {\it The condensate $\langle{\cal
O}\rangle$ as a function of $\mu L$. We set $\nu=1.2$, $Z_A(\phi)=e^{\gamma\phi}$ and $Z_\psi=e^{\alpha \phi}$. We choose  $T\simeq 0.15/L$ while  $T\simeq 0.010/L$ in the inset.}}}
\label{phase-transition-SC-?}
\end{figure}

The field equations for the  ansatz in (\ref{ansatz}) are
\bea \frac{1}{W} \partial_z(W f Z_{\psi}(\phi) \partial_z\psi) +\frac{Z_{\psi}(\phi)}{f} A_0^2 \psi =0,
\qquad \partial_z(Z_A(\phi) \partial_zA_0)-\frac{2W Z_\psi(\phi)}{L^2f} \psi^2 A_0=0. \label{field-equations}\eea  
we can now see that $A_0$ is needed: setting $A_0=0$ would imply either $\psi=0$ or  $\psi(z)\propto \int_0^{z} dz'/(W f Z_{\psi})$, which is singular at the black hole horizon. The requirement of regularity gives the following boundary conditions at $z_h$ 
\be \partial_z\psi =0, \qquad  A_0=0.  \label{IR}\ee
The condition $A_0=0$ ensures that $A_\alpha$ has finite norm \cite{Hartnoll:2008vx}. 
The second equation in (\ref{field-equations}) and the regularity of the solution implies that $A_0$ should 
go to zero at least as fast as $z_h-z$ and then the second term in the first equation of (\ref{field-equations})
should vanish at $z=z_h$. The  remainder of that first equation is a term proportional to $f$, 
which goes to zero for regular solutions, and one proportional
 to $f'$ that can be easily simplified to obtain the first condition in Eq. (\ref{IR}).

We have solved numerically the field equations in (\ref{field-equations}) with the UV and IR boundary conditions, in (\ref{UV}) and (\ref{IR}) respectively. In Fig. \ref{phase-transition-SC-?} we give the condensate as a function of the chemical potential.

Like in \cite{Hartnoll:2008vx} there is a  critical chemical
potential $\mu_c$ above which $\mathcal{O}$ condenses, but here $\mu_c/T$ depends on $TL$, $\nu$ and the functional form of $Z_{A}$ and $Z_{\psi}$. Inverting this relation one has the critical temperature of the superfluid phase transition, $T_c$. We have 
that $T_c/\mu$ is  a function
of $\mu L$, $\nu$, $Z_{A}$ and $Z_{\psi}$.
By computing explicitly the free energy, Eq. (\ref{free-energy}), one can see that, when it exists,
the superconducting solution is energetically favorable with respect to the normal one.

We observe that the condensate in Fig. \ref{phase-transition-SC-?} seems to diverge as $\mu L\rightarrow \infty$. 
Since increasing $\mu L$ for $TL$ fixed corresponds to exploring the region of small $T/T_c$, 
this suggests that $L^3 \langle{\cal
O}\rangle$ becomes bigger and bigger by decreasing $T/T_c$. We notice that neglecting the backreaction
of the charged scalar and the gauge field on the metric
is still consistent in the strict $G_N\rightarrow 0$ limit, which we are using throughout this paper\footnote{We 
plan to go beyond the $G_N\rightarrow 0$ limit in a future publication.}.

\subsection{Conductivity in the broken phase}

We now compute the conductivity in the spontaneously broken phase, $\langle \mathcal{O} \rangle \neq 0$. Most of the equations given in Section \ref{Conductivity-section} are still valid here, but some of them have to be modified because the order parameter is nonvanishing:
the linearized Maxwell equation, Eq. (\ref{linearizedEOM}), now becomes
\be  \partial_z(f Z_A(\phi) \partial_zA_x)+ \omega^2 \frac{Z_A(\phi)}{f} A_x-\frac{2 W}{L^2} Z_\psi \psi^2 A_x= 0. \ee
In particular  the equation for the modulus of the EM wave, parametrized by $s$, is modified to 
\be \partial_z(fZ_A\partial_zs) +fZ_A (\partial_zs)^2+\omega^2 \frac{Z_A}{f} \left(1-\frac{e^{-4s}}{Z_A^2}\right)-\frac{2 W}{L^2} Z_\psi \psi^2 =0 \label{s-eq-SP}\ee 
and has to be solved with the regularity boundary conditions
\be s=-\frac{1}{2}\ln Z_A, \qquad s'=\frac{2f'W Z_\psi \psi^2 /L^2 -2\omega^2 Z_A'}{Z_A(4\omega^2+f'^2)}, \qquad \mbox{at } \quad z=z_h.  \label{s-bc-SP}\ee

\begin{figure}[t]
   \begin{tabular}{cc}
   {\hspace{-0.5cm}}
   \includegraphics[scale=0.8]{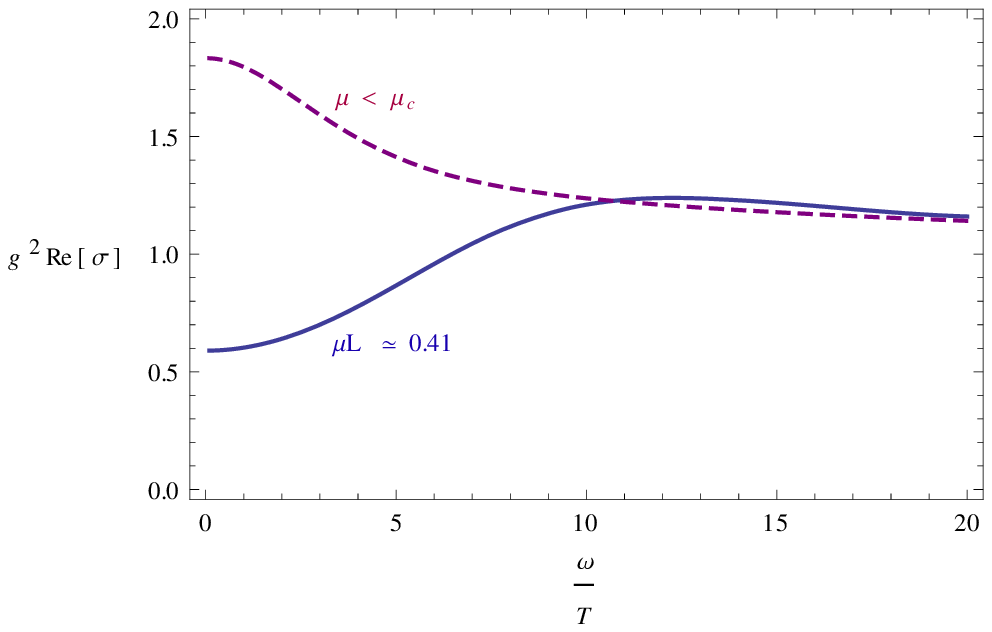}  
 
     {\hspace{1.cm}} 
    \includegraphics[scale=0.8]{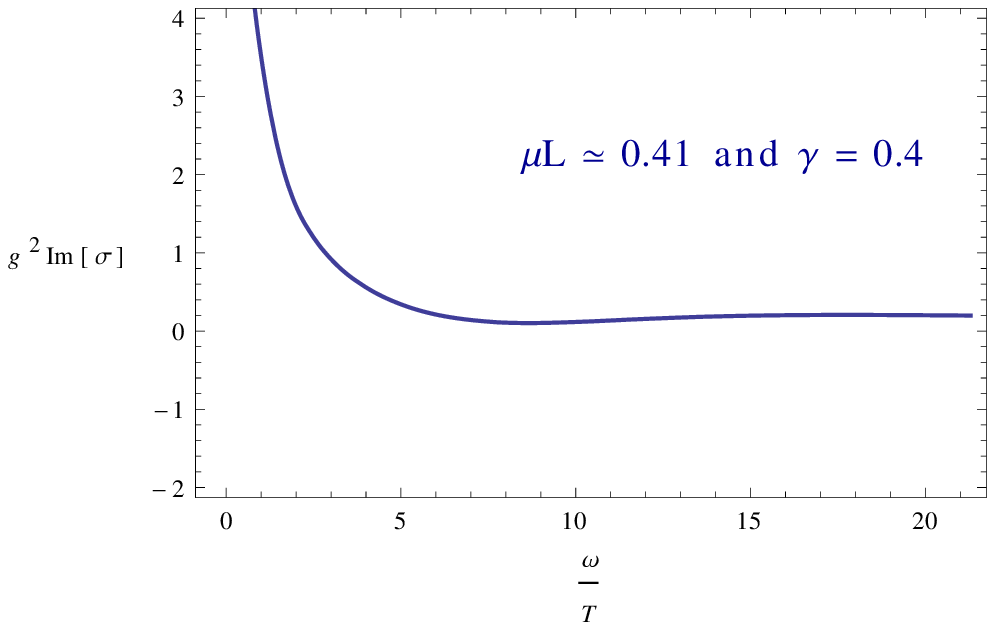}  
    \end{tabular}
\begin{tabular}{cc}
   {\hspace{-0.5cm}}
   \includegraphics[scale=0.8]{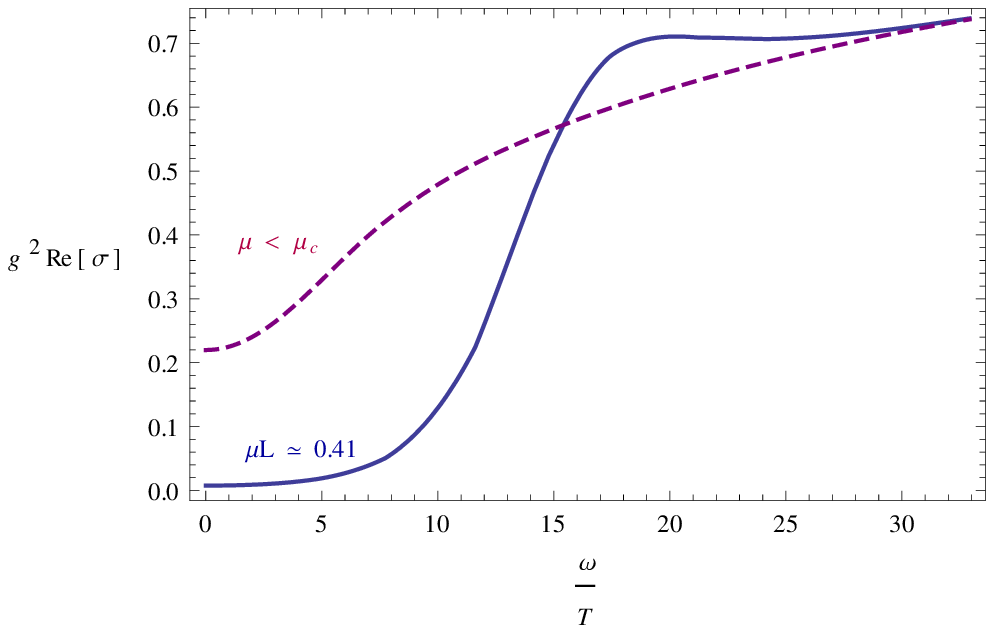}  
 
     {\hspace{1.cm}} 
    \includegraphics[scale=0.8]{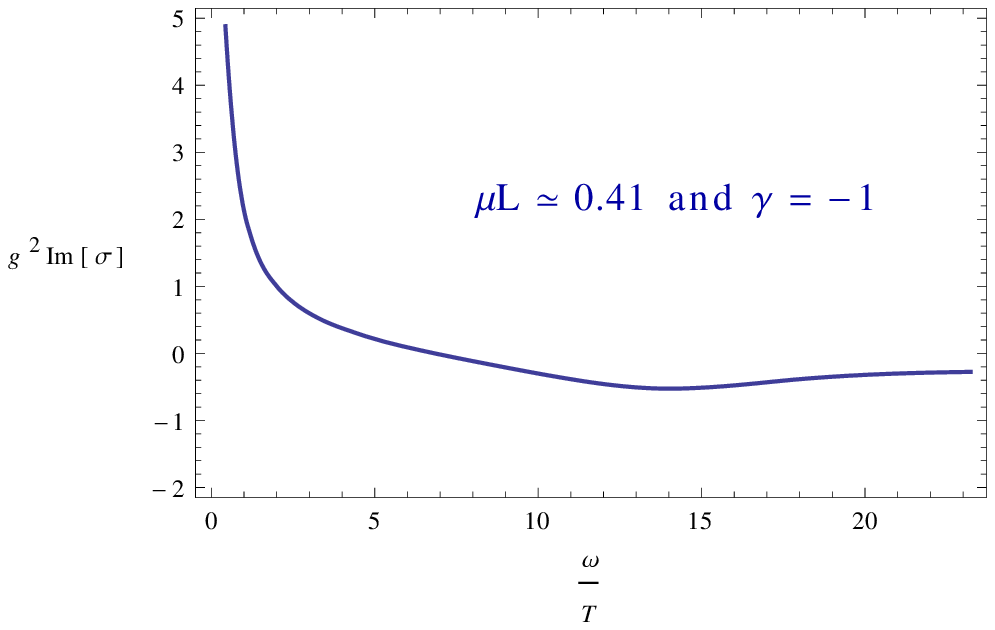}  
    \end{tabular}
   \caption{{\small {\it The conductivity as a function of the frequency  in the broken phase, $\mu> \mu_c$, (solid blue lines) and in the unbroken phase (dashed purple line). $\mbox{Re}[\sigma]$ has also a term  $\propto \delta(\omega)$.  We set $T\simeq 0.010/L$, $\nu=1.2$, $Z_A(\phi)=e^{\gamma\phi}$ and $Z_\psi=1$.
The upper plots have $\gamma=0.4$ (for which  $\mu_c\simeq 0.37/L$)  while the lower ones correspond to $\gamma=-1$ ($\mu_c\simeq 0.26/L$).}}}
\label{sigma-SP}
\end{figure} 

At very high frequencies, Eqs. (\ref{s-eq-SP}) and (\ref{s-bc-SP}) indicate that the conductivity should converge to the value in the normal S-BH phase, regardless of the presence of the condensate. This corresponds to the physical fact that at energies
much bigger than the condensation scale the system behaves as there were no condensate at all. For lower values of $\omega$ 
we provide $\sigma$ 
in Fig. \ref{sigma-SP}. The imaginary part diverges like $1/\omega$  as $\omega\rightarrow 0$, which corresponds, through the  Kramers-Kronig relation
\be \mbox{Im}[\sigma(\omega)] = -\frac{1}{\pi} \mathcal{P} \int_{-\infty}^\infty d\omega'\frac{\mbox{Re}[\sigma(\omega')]}{\omega' -\omega },\ee
to a delta function in the real part, Re$[\sigma(\omega)]\sim \pi n_s \delta(\omega)$. This means that the system is superconducting, or a superfluid in the fluid mechanical case.  The quantity $n_s$ can be identified with the superfluid density. We also see that the presence of the condensate leads to a superconducting gap in the plot of Re$[\sigma]$ versus $\omega$, like in the S-BH  phase \cite{Hartnoll:2008vx}.
In the dilaton phase, however, the gap depends on $Z_A(\phi)$.
From the Im$[\sigma(\omega)]$ plot we notice that the superfluid density is also changed by such function of the dilaton.

\section{Holographic superconductors in dilaton-gravity}\label{HSCs section}

As we mentioned in the introduction, the distinctive  property of superconductors  is that the U(1) symmetry 
is gauged rather than global, as opposed to the superfluid case; 
all we have discussed so far can be applied equally well to superconductors and superfluids because 
the dynamics of the gauge field was not important\footnote{This statement refers to physical phenomena
described by $a_\mu$ and the condensate (a U(1) breaking field) only:
%, which do not require 
%operators corresponding to the condensate constituents
when we say that the gauge field is (is not) dynamical we mean that the backreaction of the condensate on the gauge 
field 
has (has not) been taken into account, but we do not state anything about the interaction between $a_\mu$ and the condensate constituents (whose operators are not explicitly introduced here).}. The purpose of this section is to 
consider field configurations (mostly vortices) where the dynamics of the gauge field is instead crucial. When the gravitational theory is described by Einstein's gravity, this has been done in 
\cite{Domenech:2010nf}; here we will consider the case of dilaton-gravity.

\subsection{Dynamical gauge fields in AdS/CFT} 

In the standard AdS/CFT correspondence the boundary gauge field is considered as an external source for the U(1) current and as such is a nondynamical field, introduced through a Dirichlet boundary condition
(the second equation in (\ref{s-amu})). However, it is possible to promote $a_{\mu}$ to a dynamical field by substituting this Dirichlet boundary condition with one of the Neumann type, which for 
a 2+1 dimensional CFT is
\be \frac{1}{g^2}{\cal F}_z^{\,\,\, \mu} \Big|_{z=0} +\frac{1}{e_b^2}\partial_\nu {\cal F}^{\nu \mu}\Big|_{z=0}+J^\mu_{ext}=0,
\label{maxwell2} \ee
where $J_{ext}^{\mu}$ is an external nondynamical current and we have added a bare boundary kinetic term $\frac{1}{e_b^2}\partial_\nu {\cal F}^{\nu \mu}\Big|_{z=0}$ for generality \cite{Domenech:2010nf}.
This condition requires to add the following boundary terms to the action in (\ref{action})
\be
\int d^{4} x \left[
-\frac{1}{4e_b^2}{\cal F}_{\mu\nu}^2+A_\mu J_{ext}^\mu\right]_{z=0}.
\label{extrat}
\ee
In the 2+1 dimensional case, which we are considering, the bare kinetic term in (\ref{maxwell2}) is not necessary \cite{Witten:2003ya} and one can preserve conformal symmetry in the UV.
 In the absence of such term the dynamical $a_{\mu}$ can be considered as an
emergent gauge field \cite{Domenech:2010nf}. 

This method to introduce a dynamical gauge field has been originally applied to the holographic superconductor model \cite{Hartnoll:2008vx} (including diverse dimensions \cite{Horowitz:2008bn})
 in \cite{Domenech:2010nf} and we refer to this paper for further
details. It has been subsequently  applied to holographic p-wave superconductors \cite{Gubser:2008wv}-\cite{Ammon:2009fe} in \cite{Murray:2011gr,Gao:2012yw}. It can be applied without any modification to the 
dilaton-gravity model of holographic superconductors  presented here because the BH solutions we consider are all asymptotically AdS.

\subsection{Vortex phase}

Let us now look for genuine superconductor vortices with this recipe in mind. The simplest ansatz for the bulk fields is \cite{Albash:2009ix,Montull:2009fe,Albash:2009iq}
\be\Psi=\psi(z,r)e^{i  n\phi} ,\quad A_0=A_0(z,r), \quad
A_{\varphi}=A_{\varphi}(z,r)  \label{ansatz-vortex}   \ee
and the other components of $A_\alpha$ set equal to zero. In Eq. (\ref{ansatz-vortex}) $r$ and $\varphi$ are the polar coordinates, $dy^2=dr^2+r^2d\varphi^2$, restricted to $0\leq r \leq \infty$ and $0\leq \varphi <2\pi$, and $n$ is an integer. 
The corresponding field equations are
\bea \frac{1}{W} \partial_z(W f Z_{\psi}(\phi) \partial_z\psi) + \frac{Z_{\psi}(\phi)}{r}\partial_r(r\partial_r \psi)+Z_{\psi}(\phi)\left(\frac{A_0^2}{f} -\frac{(A_{\varphi}-n)^2}{r^2} \right) \psi 
 &=& 0, \nonumber \\
\partial_z\left(f Z_A(\phi)\partial_z A_{\varphi}\right)+Z_A(\phi)r \partial_r\left(\frac{1}{r} \partial_r A_\varphi\right) - \frac{2}{L^2} W Z_{\psi}(\phi)  \psi^2 (A_\varphi-n)  &=& 0,\label{eom-vortices} \\
\partial_z \left(Z_A(\phi)\partial_zA_0\right)+
\frac{Z_A(\phi)}{rf} \, \partial_r\left( r \partial_r A_0\right)- \frac{2 W Z_\psi(\phi)\,
}{L^2 \, f} \psi^2 A_0 \,  &=& 0 . \nonumber\eea
Requiring regularity we obtain the following boundary conditions at $z=z_h$ (with a 
reasoning similar to that below Eq. (\ref{IR}))
 \bea f'(z_h)\partial_z \psi +
\frac{1}{r}\partial_r(r\partial_r \psi) -\frac{(A_{\varphi}-n)^2}{r^2}
\psi &=& 0,
 \nonumber \\
 f'(z_h)\partial_z A_{\varphi}+r \partial_r\left(\frac{1}{r} \partial_r A_\varphi\right) - \frac{2 W Z_\psi(\phi)}{L^2 Z_A(\phi)}\psi^2 (A_\varphi-n)  &=& 0,\label{bch}\\
A_0&=&0,\nonumber\eea 
and at $r=0$
% %
\bea
 \partial_r A_0&=&0\,\qquad   A_{\varphi} =0\,  ,\nonumber\\
  \partial_r \psi&=&0 \ \  \ \mbox{for}\ n=0,\qquad   \psi=0 \ \ \ \mbox{for}\ n\neq 0\, .
  \label{bcr0}
  \eea
For $r\rightarrow \infty$ we impose instead the physical condition that the fields should approach the homogeneous superconducting configuration:
\be \partial_r\psi=0\ ,\ \ \  \partial_r A_0=0\ ,\  \ \  A_{\varphi}=n
\label{rinftbc}.\ee
Finally let us consider the boundary conditions at $z=0$. Requiring the U(1) to be spontaneously broken we impose (\ref{UV}), while demanding the magnetic field $B=\partial_r a_{\varphi}/r$ to be dynamical we have, from Eq. (\ref{maxwell2}),
\be
\frac{1}{g^2}\partial_z A_\varphi\Big|_{z=0} +\frac{1}{e_b^2}r\partial_r\left(\frac{1}{r}\partial_r  A_\varphi\right)\Big|_{z=0}=0\label{newbcsc}\, ,
\ee
where we have set $J^\varphi_{ext}=0$.

\begin{figure}[ht]
   \begin{tabular}{cc}
   % {\hspace{-0.5cm}}
  \includegraphics[scale=0.75]{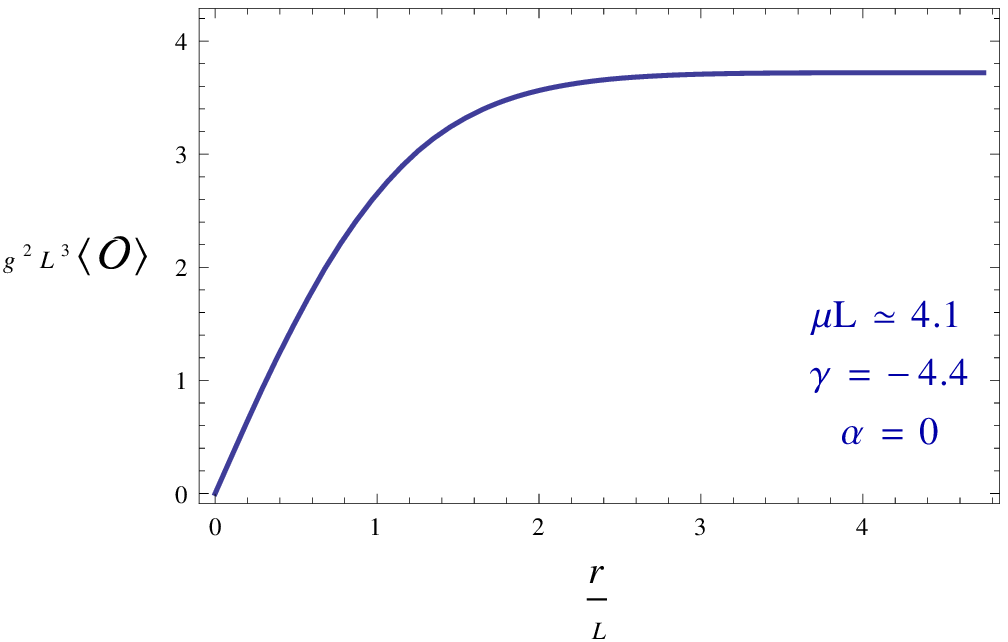}  
 
    {\hspace{1.cm}} 
    \includegraphics[scale=0.72]{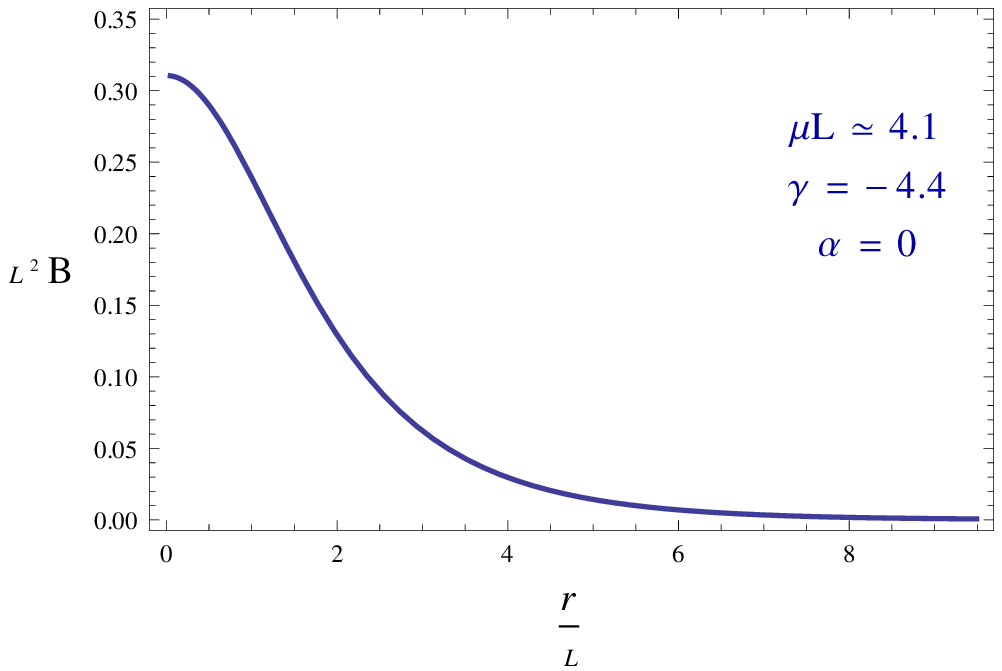}  
   \end{tabular}
 \begin{tabular}{cc}
   % {\hspace{-0.5cm}}
  \includegraphics[scale=0.75]{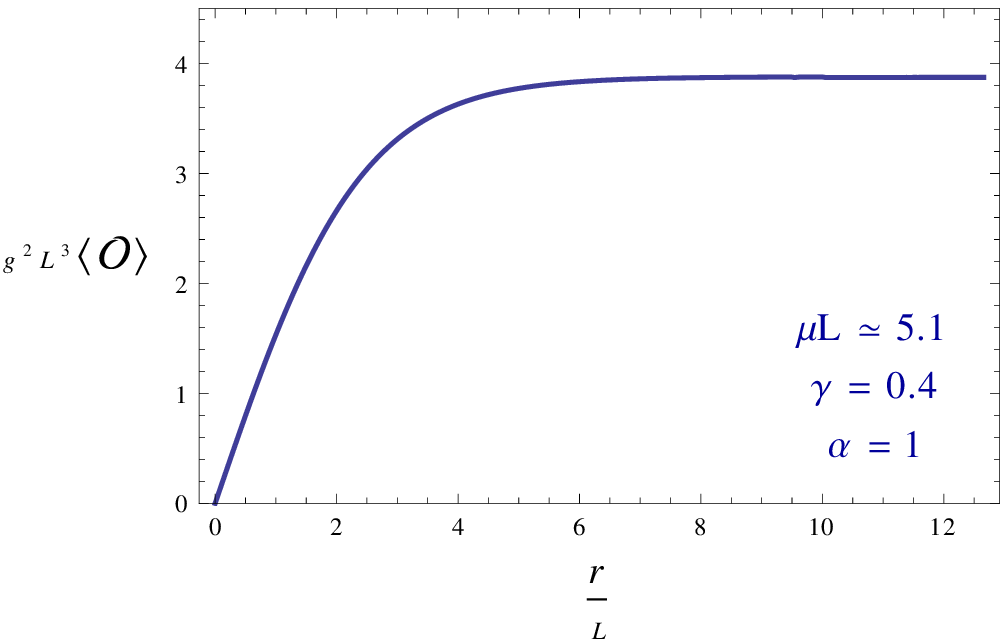}  
 
    {\hspace{1.cm}} 
    \includegraphics[scale=0.72]{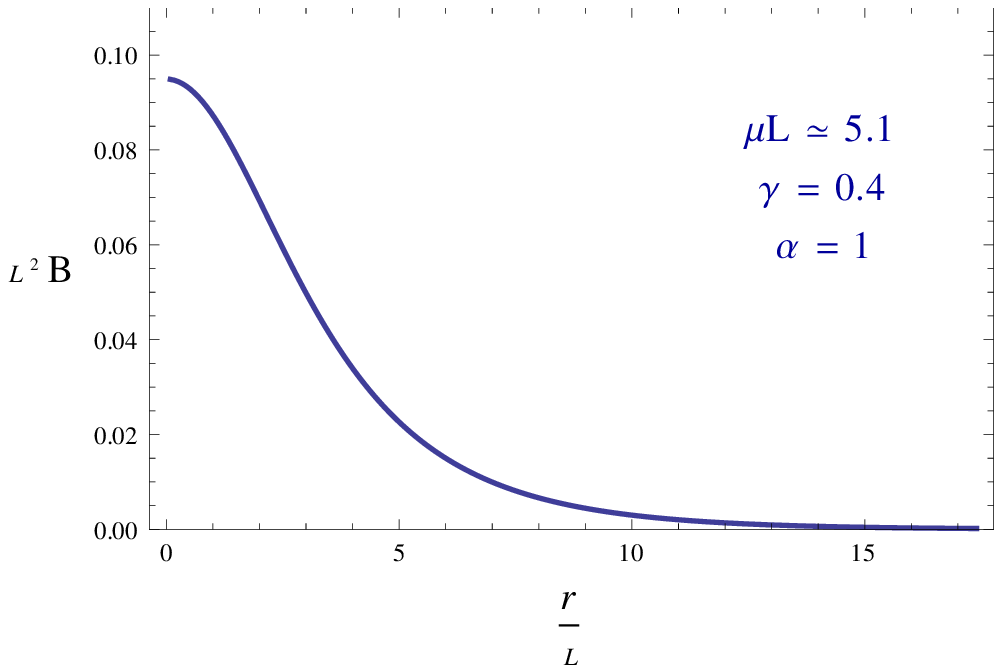}  
   \end{tabular}
   \caption{{\small {\it The condensate (left plots) and the magnetic field (right plots) for the $n=1$ vortex. Here we set $\nu=1.2$, $T\simeq 0.15/L, Z_A(\phi)=e^{\gamma \phi}$, $Z_\psi(\phi)=e^{\alpha \phi}$
and the kinetic boundary term in (\ref{maxwell2}) equal to zero, $g/e_b\rightarrow 0$.}}}
\label{OBvsr}
\end{figure}

Imposing these boundary conditions we have solved numerically  Eqs. (\ref{eom-vortices})  by using the COMSOL package \cite{comsol}. Solutions of this form correspond to straight vortex lines centered at $r=0$.
For $n=0$ the only solution we find is the homogeneous solution of Section \ref{homogeneousSFphase} with $B=0$: this is the famous Meissner effect of superconductors.
In Fig. \ref{OBvsr} we give the condensate and the magnetic field as a function of $r$ for the $n=1$ vortex.
We have checked that our solutions respect the model-independent properties given in Table \ref{table}.

\begin{figure}[ht]
   \begin{tabular}{cc}
    {\hspace{0.5cm}}
  \includegraphics[scale=0.6]{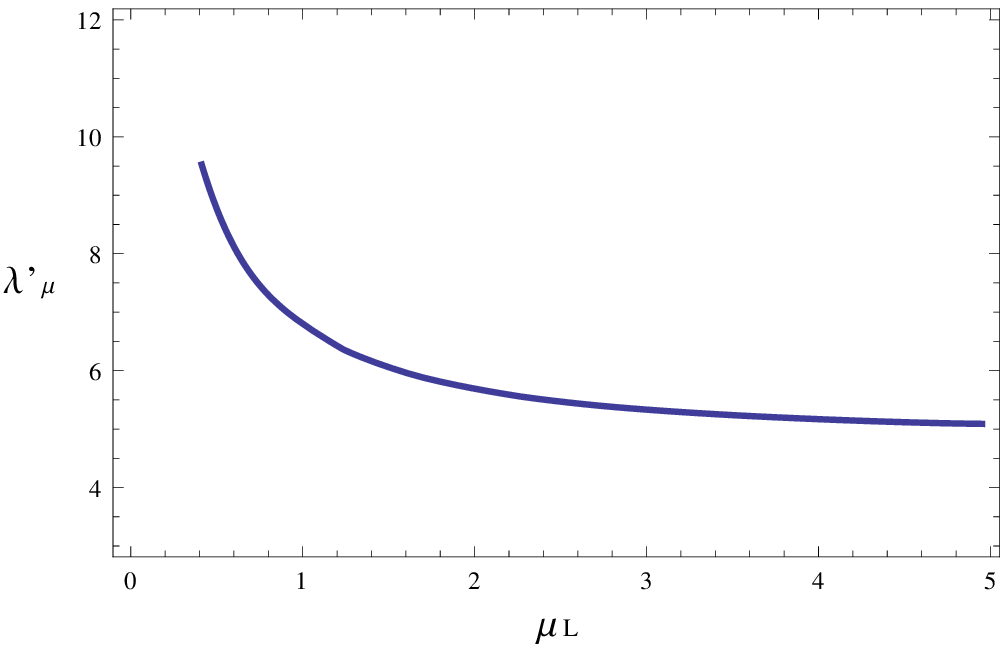}  
 
    {\hspace{2.cm}} 
    \includegraphics[scale=0.6]{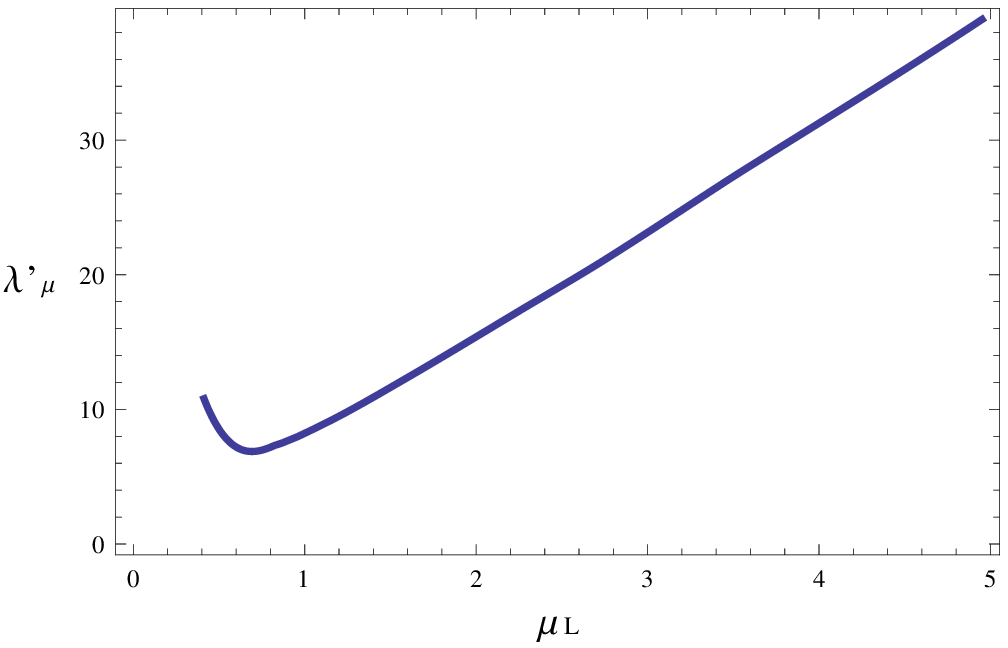}  
   \end{tabular}
   \caption{{\small {\it The penetration depth as a function of $\mu L$. Here we set $\nu=1.2$, $T\simeq 0.010/L$, $Z_A(\phi)=e^{\gamma \,\phi}$,  $Z_\psi(\phi)=1$ and $g/e_b\rightarrow 0$.
 We choose $\gamma=-4.4$ (for which $\mu_c\simeq 0.17/L$) on the left and $\gamma=0.4$  (for which $\mu_c\simeq 0.37/L$) on the right.}}}
\label{penetration-depth-vs-mu}
\end{figure}

Then we have studied the prediction of our model for the penetration depth $\lambda'$ and we present it in Fig. \ref{penetration-depth-vs-mu}. This constant characterizes, among other things, the behavior of $a_\varphi$ far away from the center of the vortex 
(see Table \ref{table}) and this allows us 
to extract it from our solutions. Let us take $Z_A(\phi)= e^{\gamma \phi}$ for the sake of definiteness. We observe that when $\gamma$ is negative enough $\lambda'$ has a qualitatively different large $\mu$ behavior with respect to the S-BH phase (and to less negative or positive values
of $\gamma$). Since increasing $\mu L$ for a fixed value of $TL$ corresponds to exploring the region of small $T/T_c$, we conclude that the superconductors in the dilaton-BH phase have a qualitatively different low temperature penetration depth.
 In particular $\lambda'$ does not increase its value more and more as  $T/T_c$ is decreased,
 contrary to what happens in the S-BH phase \cite{Domenech:2010nf}. In the low $\mu$ region we eventually find that $\lambda'$ diverges at $\mu=\mu_c$: this is true in any model of superconductivity because when $\mu \simeq \mu_c$ the condensate becomes small and
 the Ginzburg-Landau theory can be applied to predict that divergence.

\begin{figure}[ht]
   \begin{tabular}{cc}
    {\hspace{0.5cm}}
  \includegraphics[scale=0.6]{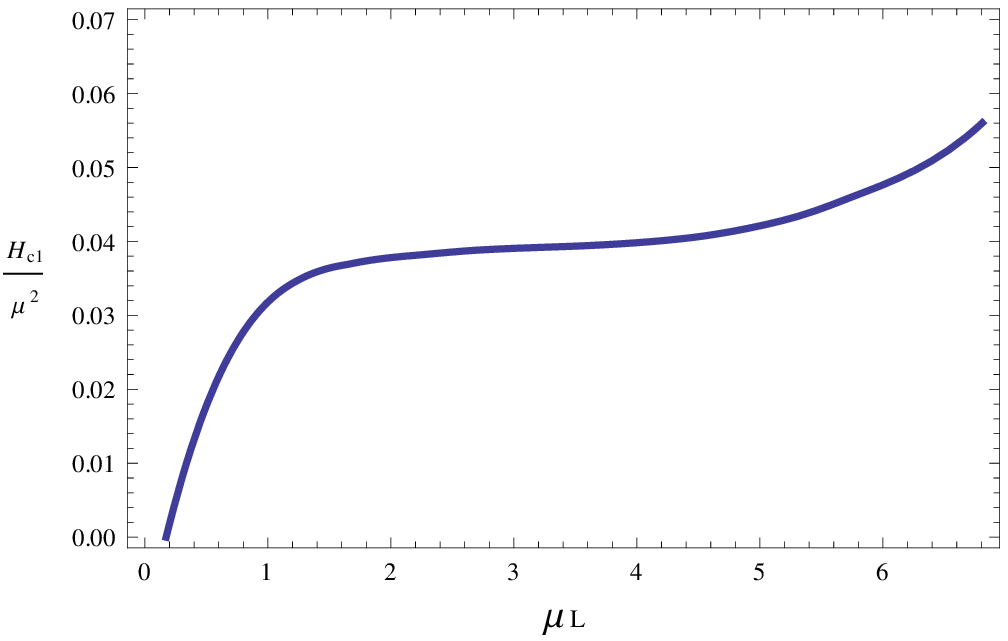}  
 
    {\hspace{2.cm}} 
    \includegraphics[scale=0.6]{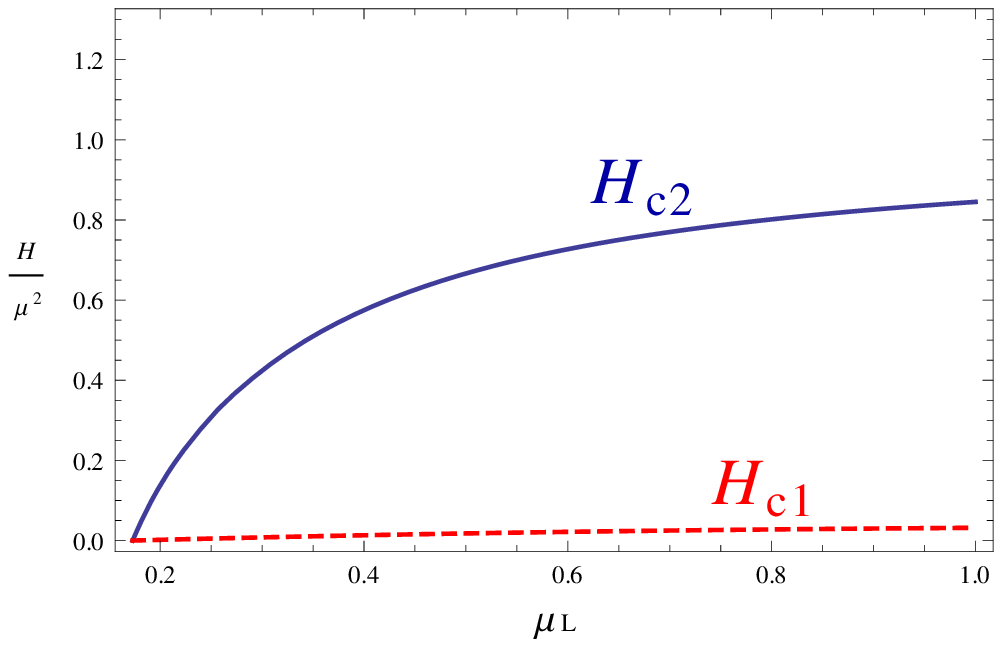}  
   \end{tabular}
   \caption{{\small {\it Left plot: $H_{c1}$ as a function of $\mu L$. Right plot: comparison between $H_{c2}$  and $H_{c1}$. Here we set $\nu=1.2$, $T\simeq 0.010/L$, $Z_A(\phi)=e^{\gamma \,\phi}$, $\gamma=-4.4$,  $Z_\psi(\phi)=1$ and $g/e_b\rightarrow 0$.
 }}}
\label{Hc12-vs-mu}
\end{figure}

Moreover, we have extracted the prediction for the first critical field $H_{c1}$, the minimum value of the external magnetic field $H$ for which the system is in the  vortex phase. This quantity can be computed with the text book formula given in Table \ref{table}.
There $e_0$ is the U(1) gauge constant of the CFT in the normal phase, $\langle \mathcal{O}\rangle=0$, normalized in a way that the kinetic term of a stationary vector potential in the  free energy is $\int d^2x \mathcal{F}_{ij}^2/(4e_0^2)$;
taking $ \mathcal{F}_{ij}$ to be constant we obtain
\be \frac{1}{e_0^2}=  \frac{1}{g^2}\int_0^{z_h} dz Z_A(\phi). \label{e0}\ee
We provide $H_{c1}$ as a function of $\mu$ in the left plot of  Fig. \ref{Hc12-vs-mu}. In the high $\mu/\mu_c$ (low $T/T_c$) region we find, like for the penetration depth, substantial qualitative differences compared to the S-BH case if the exponent $\gamma$
 is negative enough: while for the S-BH (and for less negative or positive values of $\gamma$)  $H_{c1}\rightarrow 0$ \cite{Domenech:2010nf}, in the other cases $H_{c1}$ remains sizable even at low temperatures.
The reason why this occurs is because $H_{c1}\propto e_0^2$, whose low temperature behavior depends strongly on whether we are in the scale invariant phase, $\phi=0$, or in the phase in which scale invariance is broken, $\phi\neq 0$. Eq. (\ref{e0}) tells us that in the scale invariant phase $e_0 \rightarrow 0$ and so does $H_{c1}$, but when scale invariance is broken $e_0$ remains finite if $\gamma$ is negative enough. From Eq. (\ref{dilaton-BH}) 
we see that the maximum value of $\gamma$
for which this is true is $-\sqrt{2/(\nu^2-1)}$. Interestingly, these negative values of $\gamma$ lead to an insulator at low temperatures in the normal phase. It is possible to reformulate this discussion in more general terms than exponential
$Z_A(\phi)$: what actually matters here is whether  the quantity $Z_A(\phi)|_{z=z_h}$ is small enough for small temperatures; comparing with Eq. (\ref{DCsigma}),  we see that when this is the case the normal phase tends to be insulating.
In the region where $\mu \simeq \mu_c$ we have that $H_{c1}$ decreases and eventually vanishes at $\mu=\mu_c$: this is a model-independent feature of superconductivity because at the critical point both the superconductive and the vortex configurations
go to the normal one.

 Finally, we have analyzed the second critical field $H_{c2}$, defined as the maximum value of $H$ for which we are in the vortex phase. Since $\Psi $ goes to zero continuously as $H\rightarrow H_{c2}$, the backreaction of $\Psi$ on $B$ is so small for $H\simeq H_{c2}$
that one can take $H=B$ and neglect the dynamics of the magnetic field when computing $H_{c2}$. This means that we can set a Dirichlet condition for $A_i$ at the AdS-boundary and  that $H_{c2}$ is the same for superconductors and superfluids. We will describe explicitly  how we computed this quantity in   
Section \ref{superlfuids}, but here we present the final result. In Fig. \ref{Hc12-vs-mu} we give $H_{c2}$ and we compare it with $H_{c1}$. We observe that $H_{c2}$ is always bigger than $H_{c1}$ and therefore there is a finite range of $H$, that is $H_{c1} < H< H_{c2}$,
for which the superconductor is in the vortex phase (the superconductor is of Type II). The  holographic superconductor in dilaton-gravity shares this feature with the other holographic superconductors studied so far,  based on 
pure Einstein gravity, both in ungapped \cite{Domenech:2010nf}  and gapped  \cite{Montull:2012fy} phases, and therefore it seems, at least to some extent, a universal property of holographic superconductivity.  

Since we are dealing with a Type II superconductor we can  conclude that for $H>H_{c1}$, but still close enough to $H_{c1}$, the energetically favorable configuration is the single vortex with $n=1$, while if $H$ increases further, 
more and more vortices appear until they form a triangular lattice  
 at $H \simeq H_{c2}$ \cite{triangular}.

\begin{table}[t]
\begin{center}
\begin{tabular}{|l|l|l|}
\hline  & & \\  & {\bf Superfluid (SF)}  & {\bf Superconductor (SC)}  \\ 
 & & \\  & & \\ 
 {\it {\color{greeny}$\langle J^i \rangle$ } }& SF current density  &  EM current density  \\
& & \\ 
{\color{greeny}arg{\it$(\mathcal{O}) $} }& SF velocity potential in the lab frame  &  condensate's phase  \\
& & \\ 
 {\color{greeny}{\it$a_i$ } }& external velocity in the lab frame  &  EM vector potential  \\
& & \\ 
   {\it {\color{greeny} vortex $a_\varphi$-behavior}}& $a_\varphi$ is frozen &  $a_\varphi\stackrel{large \, r}{\simeq} n+a_1\sqrt{r}e^{-r/\lambda'}$  \\
  \phantom{asd} &\phantom{asd} &\phantom{asd}  \\
{\color{greeny}{\it quantization of $B$ in vortices } }& No  &  yes:  $\int dr \, r B =n$ \\
& & \\
   {\it {\color{greeny} vortex energy} }& $F_n-F_0\stackrel{large \, R}{\sim} n^2 \ln( R/\xi_\GL)- nBR^2/2$ &  finite as $R\rightarrow \infty$  \\  \phantom{asd} &\phantom{asd} &\phantom{asd}  \\
{\it {\color{greeny} first critical field}}& $B_{c1}\stackrel{large \, R}{\simeq}2\ln( R/\xi_\GL)/R^2\stackrel{R \rightarrow \infty}{\rightarrow }0 $  &  $H_{c 1} = e_0^2 (F_1 - F_0)/2 \pi$\\
  \phantom{asd} &\phantom{asd} &\phantom{asd}  \\
{\it {\color{greeny} second critical field}} & $H_{c2}=\frac{1}{2\xi_{\rm GL}^2}$ & $H_{c2}=\frac{1}{2\xi_{\rm GL}^2}$ 
\\
  \phantom{asd} &\phantom{asd} &\phantom{asd}  \\ \hline
\end{tabular}
\end{center}\caption{\small {\it We compare superconductors and superfluids.  $\lambda'$ is the penetration depth, a model-dependent constant, 
which is generically different from $\lambda$ (the  inverse mass of $a_i$). $a_1$ is another model-dependent constant.  $F_n$ is the free energy  of a vortex with winding number $n$, while $\xi_\GL$ is the coherence length of the Ginzburg-Landau theory.
   We denote with  $H$ the external magnetic field and we normalize it such that it coincides with the total magnetic field  $B$ at $\Psi=0$. For superfluids we have $H=B$.
See Ref. \cite{Domenech:2010nf} for a derivation of the model-independent properties in this table  using effective theory methods. }} \label{table}
\end{table}

\section{Holographic superfluids in dilaton-gravity}\label{superlfuids}

As we have already said, in the limit in which $a_{\mu}$ is nondynamical a superconductor is equivalent to a superfluid. In this section we focus on such limit.
In order not to duplicate the symbols, we use the superconductor notation here, but in Table \ref{table} we give the dictionary to interpret the various quantities in terms of superfluid physics.
In particular we are interested in the case in which an external velocity (denoted by $a_i$ in the lab frame) is performed on the system.
In the formulation of superfluidity we are considering all vectors are expressed in the frame where the external velocity is zero, unless otherwise stated.

Since superfluids are characterized by being irrotational we focus on constant external angular velocities, where the only nonvanishing component of $a_i$ is

\be a_{\varphi}= \frac{B}{2}r^2. \label{Dirichlet}\ee
The external angular velocity coincides with $B/2$. When $B$  is small enough the system is in the superfluid state that is irrotational, so the fluid angular velocity is -$a_\varphi/r^2$ and $\langle J_\varphi \rangle \neq 0$.
  At high $B$ superfluidity is eventually broken and $\langle J_\varphi \rangle =0$ (normal phase). For intermediate values of $B$ the system  is in the vortex phase. 
 To holographically describe  this situation we therefore consider once again the ansatz  in (\ref{ansatz-vortex}).

\begin{figure}[ht]
   \begin{tabular}{cc}
   % {\hspace{-0.5cm}}
  \includegraphics[scale=0.75]{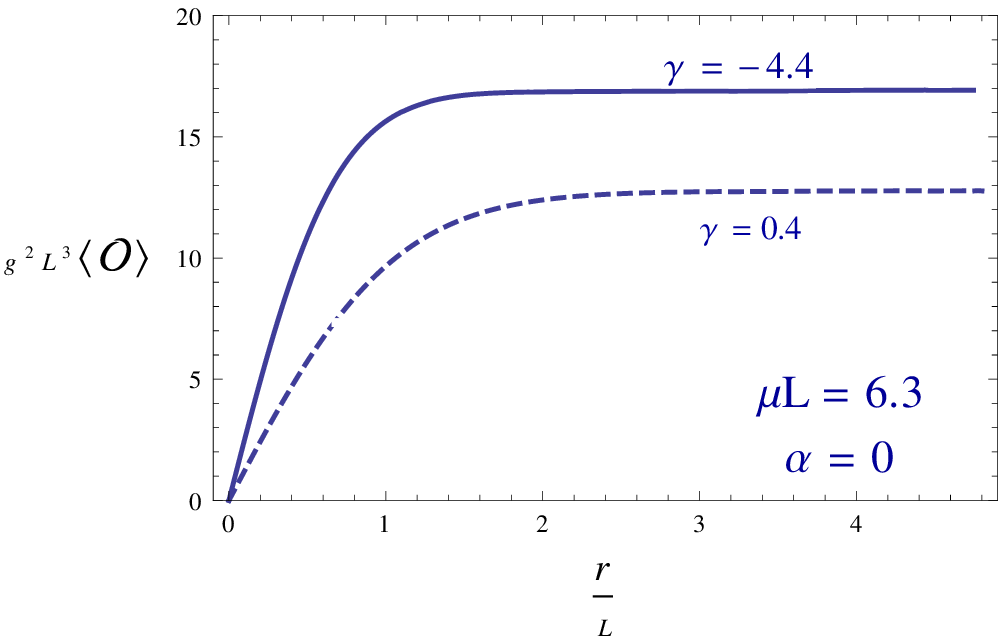}  
 
    {\hspace{1.cm}} 
    \includegraphics[scale=0.75]{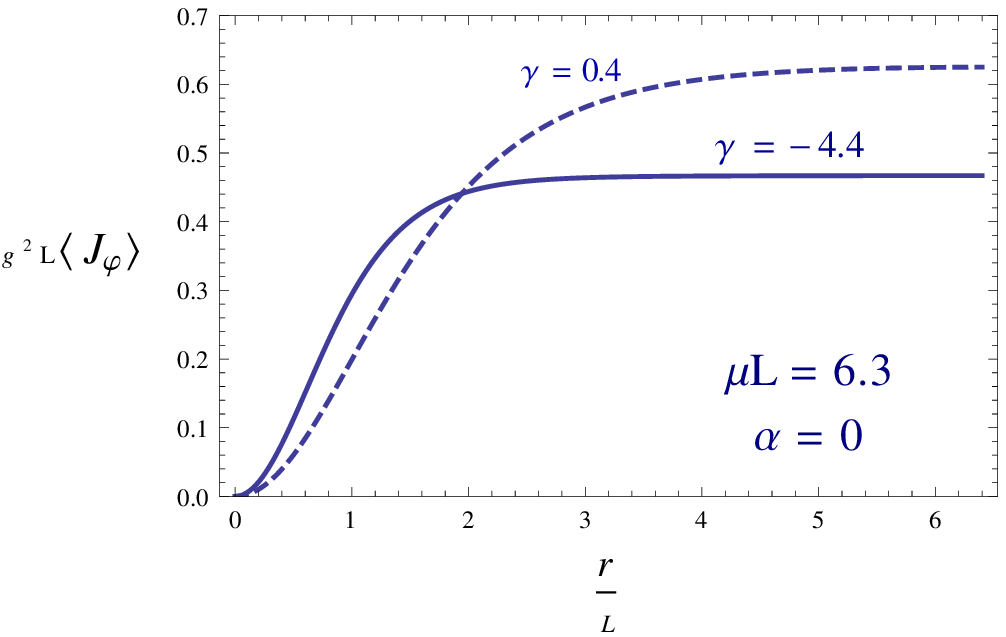}  
   \end{tabular}

\begin{tabular}{cc}
   % {\hspace{-0.5cm}}
  \includegraphics[scale=0.75]{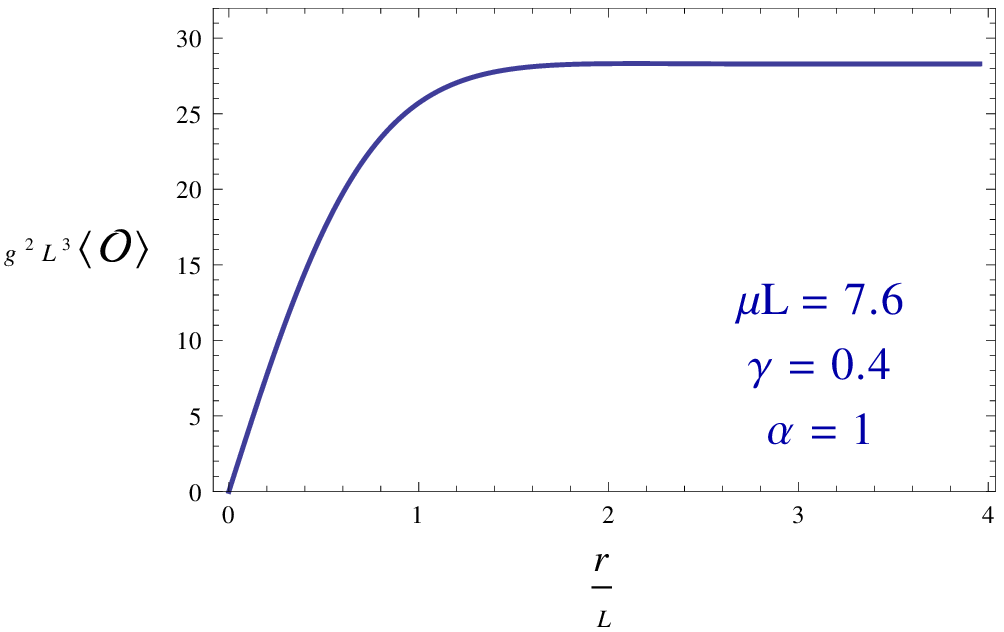}  
 
    {\hspace{1.cm}} 
    \includegraphics[scale=0.75]{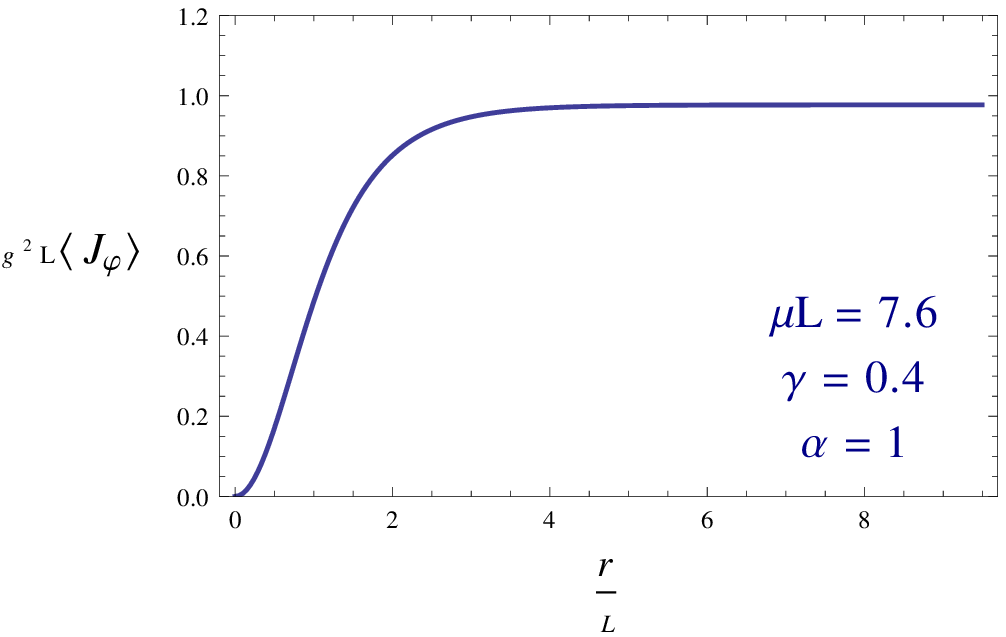}  
   \end{tabular}
   \caption{{\small {\it The condensate (left plot) and the current (right plot) for the $n=1$ superfluid vortex. In this plot we set $\nu=1.2$, $T\simeq 0.15/L$, $B\simeq 0$, $Z_A(\phi)=e^{\gamma\,\phi}$, $Z_\psi(\phi)=e^{\alpha\phi}$.}}}
\label{OJvsr}
\end{figure}

Eq. (\ref{Dirichlet})  can be viewed as a Dirichlet boundary condition at $z=0$. At large\footnote{Since the superfluid vortex free energy diverges in the infinite volume limit (see Table \ref{table}), we take a large but finite-size system: $0\leq r \leq R$. }
 values of $r$, $r=R$, we impose instead
\be \partial_r\psi=0\ ,\ \ \  \partial_r A_0=0\ ,\  \ \  A_{\varphi}=\frac{B}{2}R^2. \label{superfluid-large-r} \ee
The first two conditions correspond to the physical requirement that the field configuration goes to the homogeneous superfluid state at large $r$, while the last one  is a simple possibility compatible with the form of $a_\varphi$ given in Eq. (\ref{Dirichlet}).

We have numerically \cite{comsol} solved  Eqs. (\ref{eom-vortices}) with the boundary conditions in Eqs. (\ref{bch}), (\ref{bcr0}), (\ref{UV}), (\ref{Dirichlet}) and (\ref{superfluid-large-r}).
In Fig. \ref{OJvsr} we present the corresponding vortex profiles.

The minimum value of $B$ at which the vortex starts being favorable, called $B_{c1}$, is known for large $R$  and is given in Table \ref{table}.  That formula is model independent and so must be true here in particular. 
In the infinite volume limit, $R \rightarrow \infty$,  we have that $B_{c1} \rightarrow 0$; superfluids are deep Type II superconductors and as such they have the properties mentioned at the end of Section \ref{HSCs section}.  

As we have already stated, the maximum value of $B$ for which the system is in the vortex state, $B_{c2}\equiv H_{c2}$, coincides with that for superconductors. We have computed $H_{c2}$ as the value of $B$ at which 
$\Psi \rightarrow 0$ and provide it as a function of $\mu$ on the right plot of Fig. \ref{Hc12-vs-mu}.

\section{Outlook}\label{conclusions-section}

We have studied superconductors and superfluids through the AdS/CFT correspondence with a dilaton in the gravitational theory. Our results are summarized in the introductory section.
Let us provide some outlook of our work here.

We have focused on the case in which the charged scalar $\Psi$ is massless, this corresponds in the CFT side to an operator of dimension 3. Giving $\Psi$ a mass would change the 
dimension of $\mathcal{O}$. It may be interesting to do so as this would introduce a third function of the dilaton that represents a running mass. Along these lines one could then analyze a more general theory
with the given field content having several running constants, which may be useful to identify model-independent features.

We found that the holographic superconductors studied here are of Type II, like those without the dilaton \cite{Domenech:2010nf,Montull:2012fy} proposed in Refs. \cite{Hartnoll:2008vx,Nishioka:2009zj}. It would be valuable to know if there
is a general reason for this property or, on the other hand,  it is possible to engineer a Type I superconductor. 
A more general theory, similar to that mentioned in the paragraph above, could be a good starting point to address this question.

The (non ideal)insulator/superconductor transition that can be realized between the dilaton-BH and the corresponding condensed phase is an alternative to that proposed in Ref. \cite{Nishioka:2009zj}, which exploits a well known deconfined/confined transition 
\cite{Witten:1998zw,Horowitz:1998ha}.
It is then natural to ask if the properties of the latter scenario (see for example Refs. \cite{Montull:2011im,Bhattacharya:2012we,Montull:2012fy,Cai:2012sk}) are shared by the dilaton theories discussed here, for the setup corresponding to an insulator/superconductor transition.

We leave all these questions for future work.

\vspace{0.5cm}

{\bf Acknowledgments.} We would like to thank Andr\'es Anabal\'on for a very  useful correspondence.
This work was supported by the EU ITN ``Unification in the LHC Era", contract PITN-GA-2009-237920 (UNILHC) and by MIUR under contract 2006022501.

\end{document}